\documentstyle[12pt]{article}
\renewcommand \baselinestretch{1.4}

\begin{document}

\def\beq{\begin{equation}}
\def\eeq{\end{equation}}
\def\bce{\begin{center}}
\def\ece{\end{center}}
\def\bea{\begin{eqnarray}}
\def\eea{\end{eqnarray}}
\def\ben{\begin{enumerate}}
\def\een{\end{enumerate}}
\def\ul{\underline}
\def\ni{\noindent}
\def\nn{\nonumber}
\def\bs{\bigskip}
\def\ms{\medskip}
\def\wt{\widetilde}
\def\tr{\mbox{Tr}\, }
\def\brr{\begin{array}}
\def\err{\end{array}}

\hfill UB-ECM-PF 94/30

\hfill November 1994

\vspace*{3mm}

\begin{center}

{\LARGE \bf
Non-critical bosonic string corrections \\ to the black hole
entropy}

\vspace{8mm}

\renewcommand
\baselinestretch{0.5}
\medskip

{\sc E. Elizalde}
\footnote{E-mail: eli@zeta.ecm.ub.es} \\
Center for Advanced Study CEAB, CSIC, Cam\'{\i} de Santa
B\`arbara,
17300 Blanes
\\ and Department ECM and IFAE, Faculty of Physics,
University of Barcelona, \\ Diagonal 647, 08028 Barcelona,
Catalonia, Spain \\
 and \\
{\sc S.D. Odintsov} \footnote{E-mail: odintsov@ecm.ub.es.
On
leave from: Tomsk Pedagogical Institute, 634041 Tomsk, Russia.}
\\
Department ECM, Faculty of Physics,
University of  Barcelona, \\  Diagonal 647, 08028 Barcelona,
Catalonia,
Spain \\

\vspace{15mm}

{\bf Abstract}

\end{center}

We calculate the quantum corrections to the entropy of a very
large
black hole, coming from the theory of a $D$-dimensional,
non-critical
bosonic string. We show that, for $D >2$, as a result of modular
invariance the
entropy is ultraviolet finite, although it diverges in the
infrared
(while for
$D=2$ the entropy contains both ultraviolet and infrared
divergences).
The issue of modular invariance in field theory, in connection
with
black-hole entropy, is also investigated.

\vspace{4mm}

%\ni PACS: 04.62.+v, 04.60.-m, 02.30.+g

\newpage

The issue of black-hole entropy ---together with its puzzling
consequences--- is under permanent discussion nowadays. It is
commonly
recognized by now that the main arguments and definitions were
already
given in
the
classical works \cite{5}. However, twenty years have elapsed
since then
and many points remain still mysterious.

Focussing on our specific interest in the subject, it is well
known  \cite{13} that the black-hole entropy
calculated
in quantum field theory near the horizon diverges in the
ultraviolet.
A few months ago, some proposals have been made about the fact that
string
theory
(considered as a model for a theory of quantum gravity) might be
indeed
relevant for the description of a very large black hole
\cite{8,7}.
In particular, recently the quantum correction to the
entropy of
a large black
hole in critical string theory has been calculated in
\cite{6}. It
has been shown there that, as a result of modular invariance
(which is
associated with the gauge symmetry in string theory), the string
corrections to the entropy are ultraviolet finite ---unlike what
happens
in the field theoretical case. However, infrared divergences
appear, that are
most probably connected with the Hagedorn temperature \cite{10}
(for a
recent review on string theory at non-zero temperature see
\cite{11}).

In the present letter we extend the approach of refs.
\cite{8}--\cite{6} and
calculate the quantum correction to the entropy in a
$D$-dimensional,
non-critical bosonic string. As in the case of the critical
bosonic
string, for any $D >2$ the correction is ultraviolet finite but
it
contains infrared divergences. Then, a comparison is done with
the
corresponding
quantum corrections to the entropy that are obtained in field
theory
---this entropy
being also presented in modular invariant form.

We start with the derivation of the formula for the entropy in
the
theory of a non-critical bosonic string in $D$ dimensions.
Presumably, this will correspond
to the expression for the one-loop correction to the entropy of
a
very
large black hole in a $D$-dimensional bosonic string medium.

The expression of the one-loop free energy for a non-critical,
$D$-dimensional, bosonic string has been given in a modular
invariant
form in ref.  \cite{1}. It is the following
\bea
F(\beta) &=&- \int_{\cal F} \frac{d^2\tau}{8\pi^2\tau_2^2} \left|
\eta
\left(
e^{2\pi i \tau} \right) \right|^{-2(D-2)}
(4\pi^2\tau_2)^{-(D/2-1)} \nn
\\ && \times \sum_{n,m=-\infty}^\infty \exp \left[ -
\frac{\beta^2}{8\pi
\tau_2} (m^2 + n^2 |\tau |^2 - 2\tau_1 n m) \right],
\label{1}
\eea
where $\beta$ is the inverse temperature, $\eta (x) \equiv
x^{1/24}
\prod_{n=1}^\infty (1-x^n)$, $D$ is the physical dimension
($D=26$
corresponds to the critical bosonic string), and $\cal F$ is the
fundamental domain:
\beq
|\tau| >1, \ \ \ -1/2 < \mbox{Re} \ \tau < 1/2, \ \ \ \mbox{Im}
\
\tau >0.
\label{2}
\eeq

Notice that a very similar expression would have been obtained,
had we considered the non-critical bosonic string with the
dynamical  Weyl mode \cite{2}. However, in the explicit formula
an additional piece in the integrand appears (see for example
\cite{3}),  which depends on the radius of the compactified Weyl
mode and leads to some problems of interpretation (as that of a
continuous
string spectrum). Another remark has to do with the fundamental
region $\cal F$. If one would have performed the calculation of
the free
energy in the field theory of string modes, the integration
region in (\ref{1}) would be completely different from $\cal F$
(see nevertheless what follows). However, it is
well known \cite{4} that by doing so one erroneously overcounts
the string result and a transformation of this new region to the
fundamental one $\cal F$ must therefore be performed in any case.

Now we can turn to the calculation of the entropy ---more
precisely, of the corrections to the entropy coming from our
non-critical string. The semiclassical expression for the black
hole entropy that is widely accepted nowadays is \cite{5}
\beq
S_{BH} = \frac{A}{4\hbar G},
\label{3}
\eeq
where $G$ is the gravitational constant and $A$ is the area of
the event horizon. In order to obtain the entropy in the bosonic
non-critical string theory under consideration, we follow ref.
\cite{6} where the previous proposals of refs. \cite{7,8} have
been further developed. We first calculate the entropy density
as
\beq
S(\beta) = \beta^2 \frac{\partial F (\beta)}{\partial \beta},
\label{4}
\eeq
where $F(\beta )$ is given by (\ref{1}). Supposing now that, very
near to the horizon, the external observer can be approximated
by
a Rindler observer with a position-dependent temperature, we
substitute $\beta = 2\pi z$ to obtain the local entropy density.
Again, following ref. \cite{6} we may calculate the total entropy
as
\bea
S & =&  \int_0^\infty S(z) \, A \, dz=  \pi A \int_0^\infty  dz
\, (2\pi z)^3 \int_{\cal F} \frac{d^2\tau}{8\pi^2\tau_2^2} \left|
\eta \left(
e^{2\pi i \tau} \right) \right|^{-2(D-2)} \nn
\\ &&  (4\pi^2\tau_2)^{-D/2}  \sum_{n,m=-\infty}^\infty
(m^2 + n^2 |\tau |^2 - 2\tau_1 n m)  \exp \left[ - \frac{\pi
z^2}{2 \tau_2} (m^2 + n^2 |\tau |^2 - 2\tau_1 n m) \right].
\label{5}
\eea
Expression (\ref{5}) is modular invariant for arbitrary $D$ and,
as a result (and owing to the fact that Im $\tau =0$ is excluded from
the integration region in (\ref{5})), the entropy is ultraviolet finite
at any value of
$D>2$ (for the case of the critical bosonic string, where $D=26$, this
fact
was already noticed in ref. \cite{6}).
Again, as for the bosonic critical string, expression (\ref{5})
includes the infrared divergence produced by the tachyon. Hence,
entropy is divergent in the infrared limit. Very likely, this
issue
is closely connected with the Hagedorn phase transition \cite{10}
and,
numerically, with its associated Hagedorn temperature
\cite{10,11}, but we do not know yet how to establish this
relation explicitly.
However, we observe  that, concerning the infrared
divergence, after
properly taking care of the zero mode ($n=m=0$ in (\ref{5})),
a high-$\beta$ expansion can be performed in (\ref{5}) by
using Jacobi's theta function identity and expanding
the $\eta$ function (these techniques are described in \cite{15}).
As is easily seen from this perturbative expansion, the
expression for the free energy becomes convergent for $\beta > \beta_c
 = 2\pi \sqrt{(D-2)/3}$.
This value of $\beta_c$ is the one associated with the Hagedorn
temperature.

Expression (\ref{5}) is rather complicated, and it is quite
difficult to analyze in detail, in order to extract the full
information
that it contains. In general, only a few qualitative features can
be obtained from it, as the property of ultraviolet finiteness
mentioned
already. Nevertheless, in some special situations an analytic
treatment
is possible. For example, let us consider the case when $D=2$
(the
two-dimensional string). Then, the same calculation as in
(\ref{1}) can
be performed, what yields a very nice result, as follows
\cite{9,1}
\beq
F^{D=2}(\beta)  = \frac{1}{6\sqrt{2}\, \beta} \left(
\frac{2\pi\sqrt{2}}{\beta} + \frac{\beta}{2\pi\sqrt{2}} \right).
\label{6}
\eeq
The dual symmetry of the usual bosonic string \cite{4} is clearly
seen from (\ref{6}).

Calculating the entropy for $D=2$, one gets
\beq
S^{D=2}  = \frac{1}{3} A \ln \frac{l}{\varepsilon},
\label{7}
\eeq
where $l$ is an infrared cut-off and $\varepsilon$ an ultraviolet
one. In its structure, this result is very similar to the one
obtained in ref. \cite{7}, where in fact the same expression for
the entropy was gotten for dilatonic gravity (apart from the
factor $A$ in (\ref{7})). It must not seem strange that the
calculated correction to the entropy is divergent in both the
infrared and the ultraviolet regions, since at $D=2$ the string
is effectively reduced to a $D=2$ field theory, with the known
consequences.

We will now compare the above result (\ref{5}) for the black-hole
entropy with the corresponding result coming from $D$-dimensional
bosonic field theory. In this case the calculation of  entropy
may be performed as shown before, and leads to the following
result
\cite{6}
\beq
S(m^2) = A  \int_0^\infty  dz \, (2\pi z)^3 \int_0^\infty
\frac{ds}{s^2} (2\pi s)^{-D/2} \sum_{\tau =1}^\infty \tau^2 \exp
\left( -\frac{m^2s}{2} - \frac{2\pi^2  \tau^2z^2}{s} \right),
\label{8}
\eeq
where $m$ is the boson mass. As  was pointed out in \cite{7,8},
there is an ultraviolet divergence in (\ref{8}), since the $s$
integral grows without bound
 near $s=0$ when $z$ is small. For the massless
case and $D=2$ we can easily obtain from (\ref{8}) an exact expression
which has the same form as (\ref{7}), except for an additional
factor 1/2. In fact, for $m^2=0$,
 \beq
S (m^2=0)  =
(2\pi)^{3-D/2}  A  \int_0^\infty  dz \, z^3 \int_0^\infty
ds \,s^{-2-D/2}  \sum_{\tau =1}^\infty \tau^2 \exp
\left(  - \frac{2\pi^2  \tau^2z^2}{s} \right),
\label{71}
\eeq
and performing the change of variables $s \rightarrow t
=2\pi^2\tau^2z^2/s$ we get immediately
 \beq
S (m^2=0)  = 2^{2-D} \pi^{1-3D/2} \zeta (D) \Gamma (D/2 +1) A
\int_0^\infty dz \, z^{1-D},
\label{72}
\eeq
in particular,
 \beq
S^{D=2} (m^2=0)  = \frac{1}{6} A \ln \frac{l}{\varepsilon}.
\label{7a}
\eeq
In the general case
an explicit expression under the form of a quickly convergent series of
McDonald's functions is obtained
\beq
S(m^2) = (2\pi)^{3-D/2} A  \int_0^\infty  dz \, z^3 \sum_{\tau
=1}^\infty 2\tau^2 \left( \frac{m}{2\pi\tau  z} \right)^{D/2 +1}
K_{D/2+1} (2\pi m\tau z), \label{8a}
 \eeq
which can then be approximated and
 safely truncated after the first term \cite{15}:
\beq
K_{D/2+1} (2\pi m\tau z) \sim \left\{ \brr{ll}
\frac{1}{2} \Gamma (D/2 +1) (\pi m \tau z)^{-D/2-1}, & z \ \mbox{small},
\\ \frac{1}{2} (m \tau z)^{-1/2} \exp (-2\pi m \tau z), & z \ \mbox{not
small}. \err \right.
\label{8b}
\eeq
Substituting (\ref{8b}) into (\ref{8a}) we see that, in this massive
case and for the upper limit of the entropy integral we obtain a finite
result
\beq
\int^\infty  \sim  (2\pi)^{-(D+1)/2} \zeta (2) \Gamma ((5-D)/2) A
m^{D-2}, \label{8c}
 \eeq
in particular, for $D=2$,
\beq
\int^\infty \sim \frac{\pi A }{24 \sqrt{2}},
\label{8d}
\eeq
while in the lower integration limit, the same behaviour as for the
massless case appears
\beq
\int_0 \sim -\frac{ A }{6} \ln \varepsilon.
\label{8e}
\eeq

It is interesting to observe, on the other hand, that quantum
corrections to the entropy in field theory can be actually presented
under the form of a modular invariant quantity also. (Of course,
the
origin of SL(2,Z) invariance in  field theory has this formal
character \cite{12}, albeit not a fundamental one as in string
theory, where it is associated with  gauge symmetry itself).
Starting
from the expression for the free energy of a $D$-dimensional
massive boson
\beq
F(\beta) = -  \int_0^\infty \frac{ds}{s} (2\pi s)^{-D/2}
\sum_{\tau =1}^\infty  \exp \left( -\frac{m^2s}{2} - \frac{
\tau^2\beta^2}{2s} \right),
\label{9}
\eeq
let us formally introduce in (\ref{9}) a unity as follows
\beq
1= \sum_{k=1}^{N_0} \delta_{k1} = \sum_{k=1}^{N_0} \int_{-
1/2}^{1/2} dt \, \exp [2\pi i t (k-1)],
\label{10}
\eeq
where $N_0 >1$ is an arbitrary natural number. Then, we can
write,
\beq
F(\beta) = - \sum_{k=1}^{N_0} \int_0^\infty \frac{ds}{s} (2\pi
s)^{-D/2}  \int_{-1/2}^{1/2} dt \, \exp [2\pi i t (k-
1)]\sum_{\tau =1}^\infty  \exp \left( -\frac{m^2s}{2} - \frac{
\tau^2\beta^2}{2s} \right),
\label{11}
\eeq
and by choosing now $t=$ Re $\tau, \ s=$ Im $\tau, \ \tau =
\tau_1 +
i\tau_2$, it is easy to see that (\ref{11}) is invariant under
the change $\tau \rightarrow \tau +1$, i.e. it is invariant under
the group of discrete translations U. But, as is known, U is the
congruence (Borel subgroup) of the modular group SL(2,Z). Thus,
using the techniques of ref. \cite{12} for the field theory
propagator, we can actually rewrite (\ref{11}) in modular
invariant form:
\beq
F(\beta) = - \sum_{k=1}^{N_0} \sum_{(c,d)=1} \int_{\cal F} d^2
\tau \, \frac{\exp [2\pi i  (k-1) \mbox{Re}\, \gamma_{cd} \tau]}{
\mbox{Im}\, \gamma_{cd} \tau (2\pi \mbox{Im}\, \gamma_{cd}
\tau)^{D/2}} \sum_{\tau =1}^\infty  \exp \left( -
\frac{m^2}{2}\mbox{Im}\, \gamma_{cd} \tau - \frac{
\tau^2\beta^2}{2\mbox{Im}\, \gamma_{cd} \tau } \right),
\label{12}
\eeq
where the second sum is extended over relative prime integers $c, d$,
 $\cal F$ is again the fundamental domain, and
\beq
\gamma_{cd} = \left( \brr{cc} * & * \\ c & d \err \right),
\label{12a}
\eeq
with the $*$ denoting arbitrary elements ---in the sense that any
transformation of SL(2,Z) with the bottom row $(c,d)$ can be used
as a
representative of the coset element.

{}From here one can easily obtain, just in the same way as in
(\ref{8}),
the
entropy corresponding to the bosonic field theory in a modular invariant
form:
\bea
S &=& A \int_0^\infty dz \, (2\pi z)^3 \sum_{k=1}^{N_0}
\sum_{(c,d)=1}
\int_{\cal
F} d^2 \tau \, \frac{\exp [2\pi i  (k-1) \mbox{Re}\, \gamma_{cd}
\tau]}{
(\mbox{Im}\, \gamma_{cd} \tau)^2 (2\pi \mbox{Im}\, \gamma_{cd}
\tau)^{D/2}} \nn \\ && \hspace{1cm} \sum_{\tau =1}^\infty \tau^2
\exp
\left[ - \frac{m^2}{2}\mbox{Im}\, \gamma_{cd} \tau - \frac{
\tau^2(2\pi z)^2}{2\, \mbox{Im}\, \gamma_{cd} \tau } \right].
\label{13}
\eea
In
this manner we have been able to calculate the quantum correction to the
entropy in the bosonic field theory as a modular invariant expression,
what
contradicts the claim against this possibility
 made in ref. \cite{6}.
Although very similar in structure to (\ref{8}), expression (\ref{13})
appears to be free from the ultraviolet divergence. In fact, even if $z$
can still be zero, the reason for the ultraviolet divergence in
(\ref{8})
was the growth of the second integral (over $s$ there). This seems not
to
be the case now, since all points in the integration region $\cal F$ are
at distance $\geq 1$ from the origin of the $\tau$-plane. As we have
seen in detail before (eqs. (\ref{7a}) and (\ref{8e})), tecnically the
appearance of the singularity was drawn by a change of
variables in
the second integral that left the integration range unchanged
and introduced negative powers of
$z$ in the first, rendering it divergent. Again, this cannot be done
here without altering the $\cal F$ region. Summing up, we think that the
entropy in its modular invariant form (\ref{13}) does no exhibit the
usual ultraviolet
divergence. This should not mean that it is absolutely free from
divergences, because a possible appearance
of an new one, which could come from the sum over $c, d$ (again for
very small $z$) is not excluded without more carefull work.
Indeed, expression (\ref{13}) seems to be quite
difficult to analyze in general,
although the hope remains that it can yield workable results in some
particular cases, by using  specific zeta-function techniques
(as the ones  introduced in ref. \cite{15}).

In conclusion, we have calculated here the quantum corrections
to the
entropy for the case of a non-critical bosonic string. Such
entropy is
modular invariant in any dimension $D$ and it is ultraviolet
finite for
$D>2$. We have also found a modular invariant expression for the
entropy
in bosonic field theory. Recently, a new representation (in terms
of a
Laurent series) for the string free energy has been introduced
in ref.
\cite{14}. Quite remarkably, this representation gives a
convenient way
to perform the analytical
continuation of the free energy beyond the Hagedorn temperature.
However, modular invariance is not explicitly aparent in this
description, owing mainly to the fact that it is not an integral
representation. It would be, nevertheless, of interest to analyze
the
divergences of  string entropy in that formalism \cite{14}, what
we
are planning to do in the near future.

 \vspace{5mm}

%\newpage

\noindent{\large \bf Acknowledgments}

We  would like to thank A. Bytsenko, A. Dabholkar and S. Zerbini
for
discussions on similar problems.
EE thanks T. Muta for discussions and very kind hospitality at
Hiroshima
University.
This work has been
supported by DGICYT (Spain), project No. PB93-0035,
by CIRIT (Generalitat de Catalunya),
and by RFFR (Russia), project
No. 94-02-03234.

\end{document}